\documentclass[12pt]{iopart}
\usepackage{graphicx}
\usepackage{iopams}

\begin{document}

\article[Dark-bright solitons in two-component BECs]{}{Interaction of dark-bright solitons in two-component Bose-Einstein condensates}

\author{S.~Rajendran$^1$, P. Muruganandam$^2$ and M. Lakshmanan$^1$}

\address{$^1$Centre for Nonlinear Dynamics, Bharathidasan University, Tiruchirapalli 620 024, India}
\address{$^2$Department of Physics, Bharathidasan University, Tiruchirapalli 620 024, India}

\begin{abstract}

We study the interaction of dark-bright solitons in two component Bose-Einstein condensates by suitably tailoring the trap potential, atomic scattering length and atom gain or loss. We show that the coupled Gross-Pitaevskii (GP) equation can be mapped onto the Manakov model. An interesting class of matter wave solitons and their interaction are identified with time independent and periodically modulated trap potentials, which can be experimentally realized in two component condensates. These include periodic collapse and revival of solitons, and snake-like matter wave solitons as well as different kinds of two soliton interactions. 

\end{abstract}


\vspace{2pc}

\submitto{\JPB}

\maketitle

The advent of solitons in ultra cold quantum gases has provided a new insight into the understanding of the localized wave packets that travel over long distances without attenuation. In this context, recent explorations of solitons in Bose-Einstein condensates (BECs) have paved the way for new developments in manipulating coherent matter waves for applications, including  atom interferometry, coherent atom transport and quantum information processing or quantum computation. In particular, experimental observations of matter wave solitons of the dark~\cite{Burger1999, Denschlag2000}, and bright~\cite{Strecker2002, Khaykovich2002, Cornish2006}  types in BECs have attracted a great deal of attention in connection with the dynamics of nonlinear matter waves, including soliton propagation \cite{Busch2000, Salasnich2004}, vortex dynamics~\cite{Rosenbusch2002}, interference patterns~\cite{Liu2000} and domain walls in binary BECs~\cite{Malomed2004}.

In this connection, recently much effort has been given to the study of matter wave solitons in BECs with time varying control parameters such as (i) variation of atomic scattering length which can be achieved through Feshbach resonance~\cite{Moerdijk1995, Roberts1998, Stenger1999, Cornish2000,Courteille1998}, (ii) inclusion of appropriate time dependent gain or loss term which can be phenomenologically incorporated to account for the interaction of atomic cloud or thermal cloud, and (iii) periodic modulation of trap frequencies~\cite{Janis2005}. Bright solitons created in the experiments are themselves condensates and propagated over much larger distances than dark solitons which, on the other hand, can only exist as notches or holes within the condensates~\cite{Burger1999, Denschlag2000}. Recently, dark solitons, their oscillations and interaction have been demonstrated in experiments with single component BECs~\cite{Christoph2008, Weller2008, Stellmer2008}. Many studies on the bright/dark soliton formation and propagation of attractive/repulsive BECs have been focussed mainly on the single species systems.

Multi-component generalization of the soliton dynamics is very natural in the context of atomic BECs because of the several ways to create such systems, for example as mixtures with two different atomic species/hyperfine states~\cite{Myatt1997,Stenger1998,Papp2008}, and as internal degrees of freedom liberated under an optical trap and atom-molecule BECs~\cite{Woo2008}. The multicomponent BECs, far from being a trivial extension of the single component ones, present novel and fundamentally different scenarios for their ground states and  excitations~\cite{Esry1998,Busch1997}. Matter wave solitons in multicomponent BECs hold promise for a number of applications, including the multi-channel signals and their switching, coherent storage and processing of optical fields.

Since multi-component BECs are of greater interest for the aforesaid reasons, in this letter we investigate the interaction of dark and bright solitons in two component BECs. In the context of cold atomic gases, the two vector components which evolve under the Gross-Pitaevskii equation are the macroscopic wave functions of Bose condensed atoms in two different internal states, which we shall denote as $\vert 1\rangle$ and $\vert 2\rangle$. If one considers condensates of, for instance, $^{23}$Na and $^{87}$Rb atoms, the nonlinear interactions are due to elastic s-wave scattering among the atoms, and are effectively repulsive (positive scattering length) for both the systems in which multi-component condensates have been realized~\cite{Bongs2001}. Very recently, there has been a tremendous interest in studying the dynamics of two-component Bose-Einstein condensates coupled to the environment using both experimental and theoretical means~\cite{Syassen2008,Anglin1997, Ruostekoski1998, Vardi2001, Ponomarev2006, Wang2007}. Here we bring out exact bright-dark solitons in repulsively interacting two-component BECs coupled with the thermal clouds, which introduce gain/loss of condensate atoms. In particular, by considering both  time independent and periodically modulated trap potentials, we show fascinating interactions of matter wave solitons.

For the above purpose, we consider a trapped BEC with two components, where the dynamics takes place only in one dimension due to the strong trap confinement in the transverse direction. We have also included the interaction of external thermal clouds which is described by gain or loss term. The properties of BECs that are prepared in two hyperfine states can be described at sufficiently low temperatures by the dimensionless form of two coupled GP equations as~\cite{Busch2001}
\begin{eqnarray}
\fl i \frac{\partial \psi_{j}}{\partial t} = -\frac{1}{2}\frac{\partial^2 \psi_{j}}{\partial  x^2}  + \Biggl[ R(t) \sum_{k=1}^{2} g_{jk}\vert \psi_{k}\vert^2 \, + V(x,t) -\mu_j+i\frac{\gamma(t)}{2} \Biggr] \psi_{j}, \;\;j = 1,2. \label{cgpe}
\end{eqnarray}
Here $V(x,t)= \frac{1}{2} \Omega^2(t) x^2$ is the external potential, $\Omega^2(t)=\frac{\omega_x^2 }{\omega_{\perp}^2}$, $\omega_x$ is the trap frequency in the axial direction, $\omega_{\perp}$ is the radial trap frequency  ($\Omega^2(t) < 0$  for expulsive potential and $\Omega^2(t) > 0$  for confining potential), $\mu$ is the  chemical potential, $R(t)=\frac{2 a_s(t)}{a_B}$, $a_s(t)$ is the $s$-wave scattering length, $a_B$ is the Bohr radius and $g_{ij} = 1$  and $\gamma(t)=\frac{\Gamma(t)}{\omega_{\perp}}$, $\Gamma(t)$ is the gain/loss term, which is the phenomenologically incorporated interaction of thermal cloud ~\cite{Kohl2002,miesner1998,Gerton2000}. Here we have considered the case of a two-component condensate with $^{87}$Rb atoms prepared in two different hyperfine states for which the atomic masses, intra- and inter-component atomic scattering lengths are equal as described in Ref.~\cite{Christoph2008}. Note that in equation (\ref{cgpe}), the variable $x$ actually represents $\frac{x}{a_{\perp}}$, where $a_{\perp}=\sqrt{\frac{\hbar }{m \omega_{\perp}}}$, and similarly $t$ stands for $\omega_{\perp} t$. 

In equation~(\ref{cgpe}) when the gain/loss term $\gamma(t)$ is positive, it leads to the mechanism of loading external atoms (thermal clouds) into the BEC by optical pumping while $\gamma(t) < 0$ describes a BEC that is continuously depleted (loss) of atoms. If the condensate is fed by a surrounding thermal cloud, then the condensate undergoes an appropriate growth/collapse. In actual BEC experiments, for example, with $^7$Li and $^{85}$Rb atoms, the atomic scattering length can be varied by suitably tuning the external magnetic field through the Feshbach resonance as~\cite{Strecker2002, Khaykovich2002, Courteille1998}
\begin{eqnarray}\label{feshbach}
a_s(t)=a_s^0\left(1-\frac{\Delta}{B(t)-B_0}\right),
\end{eqnarray}
where $a_s^0$ is scattering length of condensed atoms, $B(t)$ is the external time varying magnetic field, $B_0$ is the resonance magnetic field and $\Delta$ is the resonance width. Similar tuning of scattering length should be possible for two component condensates as well, for example in the case of hyperfine states of $^{87}$Rb.

Using the following point transformation~\cite{Gurses2007, Serkin2007, Rajendran2008, Kundu2009}
\begin{eqnarray}\label{trans}
\psi_j(x,t)=\Lambda_j(x,t) \;q_j(X,T),\;\; j=1,2,
\end{eqnarray}
equation (\ref{cgpe}) can be reduced to the set of two coupled nonlinear Schr\"odinger equations (2CNLS) of the form
\begin{eqnarray}
i\frac{\partial q_j}{\partial T}=-\frac{1}{2} \frac{\partial^2 q_j}{\partial  X^2} + q_j \sum_{k=1}^{2} g_{jk} \vert q_k\vert^2, \;\; j = 1, 2, \label{2cnls}
\end{eqnarray}
where
$\Lambda_j(x,t) = r_0 \sqrt{2 \tilde{R}(t)}\exp[i (\theta(x,t)+\mu_j t)+\int \gamma(t) dt]$,  $\tilde{R}(t) = R(t) \exp\left[\int \gamma(t) dt\right]$, $\theta=-\frac{\tilde{R}_t}{2\tilde{R}}x^2 + 2 c_1 r_0^2 \tilde{R} x-2 c_1^2 r_0^4 \int \tilde{R}^2 dt$, $X=\sqrt{2} r_0 \tilde{R} x - 2 \sqrt{2} c_1 r_0^3\int \tilde{R}^2 dt$, $T= 2 r_0^2\int \tilde{R}^2 dt$ ($c_1$, $r_0$: constants), and $\tilde{R}(t)$ and $\Omega(t)$ have to satisfy the condition
\begin{eqnarray}
\frac{d}{dt}\left(\frac{\tilde{R}_t}{ \tilde{R}}\right)-\left(\frac{\tilde{R}_t}{\tilde{R}}\right)^2-\Omega^2(t)=0,  \label{riccati}
\end{eqnarray}
which is a Riccati type equation for $\frac{\tilde{R}_t}{\tilde{R}}$. Equation (\ref{2cnls}) is the so called defocussing Manakov system, which exhibits interesting one, two and $N$-soliton solutions of bright-bright, dark-dark and dark-bright types~\cite{Radhakrishnan1995, Sheppard1997, Vijayajayanthi2008}. Concentrating for the present only on dark-bright solitons, from the solutions of equation (\ref{2cnls}), one can straightforwardly construct the one, two and $N$ dark-bright soliton solutions for equation (\ref{cgpe}), provided ${R}_t$, $\gamma(t)$ and $\Omega(t)$ satisfy equation (\ref{riccati}). We may note here that because of the complicated transformation involved above, the resultant soliton parameters are no longer constants but are functions of $x$ and $t$. Therefore, the resultant soliton solutions may be considered as generalized solitons and not simple standard solitons.

\paragraph{One-soliton dynamics:-} The dark-bright components of the one-soliton solution of the defocussing Manakov system~(\ref{2cnls}) can be given as
\begin{eqnarray}\label{1bs}
\displaystyle
q_1(X,T)= & \tau \frac{1-\frac{\rho}{\rho^*} \chi \, \mbox{e}^{\eta+\eta^*}}{1+\chi \, \mbox{e}^{\eta+\eta^*}}\mbox{e}^{icX-i[c^2/2+\tau^2]T},\\
q_2(X,T)= & \frac{\mbox{e}^{\eta}}{1+ \chi \, \mbox{e}^{\eta+\eta^*}},
\end{eqnarray}
where
$\eta=\kappa X+i (\frac{1}{2}\kappa^2-\tau^2) T$,
$\kappa=a+i b$, $\rho=\kappa-i c$, $\chi=\left[(\kappa+\kappa^*)^2\left(\frac{\vert \tau \vert^2}{\rho \rho^*}-1\right)\right]^{-1}$. The parameters $a$ and $b$ correspond to the amplitude and velocity of the soliton envelope, respectively, and $c$ refers to the phase.

Now using the transformation (\ref{trans}), the corresponding dark-bright one-soliton solution of the two coupled GP equation can be written as
\begin{eqnarray}
\displaystyle
\psi_j(x,t)=&\sqrt{2 \tilde R(t)}\, q_j(X,T)\, \mbox{e}^{i \left(\theta(x,t)+\mu t\right)+\int \frac{\gamma(t)}{2} dt},\; j = 1, 2. \label{sol:dark}
\end{eqnarray}

\paragraph{Two soliton dynamics:-} The dark-bright two-soliton solution of the defocussing Manakov system~(\ref{2cnls}) can be written as
\begin{eqnarray}
\displaystyle
\fl q_1(X,T) & = & \frac{\tau}{d}\,\mbox{e}^{icX-i[c^2/2+\tau^2]T} \Biggl(1-
\sum_{j,k=1}^{2} \frac{\rho_j}{\rho_k^*} \chi_{jk} \, \mbox{e}^{\eta_j+\eta_k^*}  +\frac{\rho_1 \rho_2}{\rho_1^* \rho_2^*} f\, \mbox{e}^{\eta_1+\eta_1^*+\eta_2+\eta_2^*}\Biggr) \label{sol2:q1}\\
\fl q_2(X,T) & = & \frac{1}{d} \left(
\sum_{j,k=1}^{2} \mbox{e}^{\eta_j}- \sum_{j,k=1}^{2} \nu_{jk} \chi_{1j}\chi_{2j} \, \mbox{e}^{\eta_j+\eta_j^*+\eta_k} \right), \label{sol2:q2}
\end{eqnarray}
where
\begin{eqnarray}
d = 1+\sum_{j,k=1}^{2}  \chi_{jk}\mbox{e}^{\eta_j+\eta_k^*}+ f\, \mbox{e}^{\eta_1+\eta_1^*+\eta_2+\eta_2^*}, \nonumber
\end{eqnarray}
$\eta_j=\kappa_j X+i (\frac{1}{2}\kappa_j^2-\tau^2) T$,
$\kappa_j=a_j+i b_j$, $\rho_j= \kappa_j-i c$, $\chi_{jk}=\left[(\kappa_j+\kappa_k^*)^2\left(\frac{\vert \tau \vert^2}{\rho_j \rho_k^*}-1\right)\right]^{-1}$,
$\nu_{jk}=(\kappa_j-\kappa_k)^2\left(\frac{\vert \tau \vert^2}{\rho_j \rho_k}+1\right)$, $f=\chi_{11} \chi_{22} \vert \nu_{12} \chi_{12} \vert^2.$
The dark-bright two soliton solution of the two coupled GP equation~(\ref{cgpe}) can be again represented by equations (\ref{sol:dark}) with $q_1$ and $q_2$ of the forms (\ref{sol2:q1}) and (\ref{sol2:q2}). The procedure can be extended to $N$-soliton solutions also. However, we do not present their forms here. Depending on the form of the trap potential, gain/loss and interatomic interaction novel type of dark-bright matter wave solitons can be deduced using the above forms (\ref{sol:dark})-(\ref{sol2:q2}). In the following, we demonstrate them for two simple trap potentials. For the other choices, results will be presented elsewhere. In the present study we fixed the trap parameters similar to that used in a recent experiment on dark-bright solitons in two component $^{87}$Rb condensates~\cite{Christoph2008}.

\paragraph{ Time independent trap potential:-} First let us consider, as an example, the case of time independent expulsive parabolic trap potential, $\Omega^2(t)= -\Omega_0^2$ for which the integrablity condition (\ref{riccati}) gives $\tilde{R}(t) = \mbox{sech} (\Omega_0 t + \delta)$. 
\begin{figure}[!ht]
\begin{center}
\includegraphics[width=0.8\linewidth]{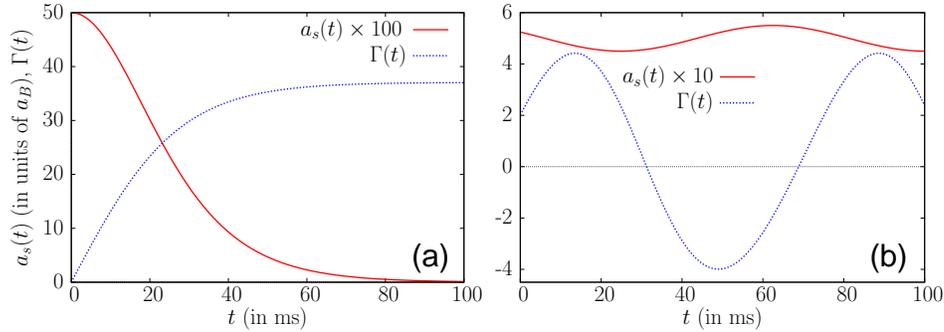}
\caption{(Color online) Choice of the atomic scattering length $a_s(t)$ and gain or loss term $\Gamma(t)$ as a function of time that exhibits dark-bright solitons for (a) time independent expulsive trap potential, $V(x) = -\frac{1}{2}\Omega_0^2 x^2$ and (b) Periodic modulated trap potential, $\displaystyle V(x,t) =  \omega^2\left\{1-\frac{4+ 10  \omega \cos(\omega t+\delta)+ 3 \omega^2 [1+\cos^2(\omega t+\delta)]}{4 \left[1+ \omega\cos(\omega t+\delta)\right]^2 }\right\} x^2$.}
\label{fig1}
\end{center}
\end{figure}
The amplitude of the wave packet is given by $\sqrt{2 R(t)} e^{\int \gamma(t)/2 dt}$. Different types of soliton solutions can be constructed for suitably chosen gain term $\gamma(t)$. In figure \ref{fig1}(a), we plot the gain $\Gamma(t) = \omega_\perp \gamma(t) = 2\pi\omega_\perp \Omega_0 \tanh(\Omega_0 t + \delta)$, and  the corresponding choice of atomic scattering length $a_s(t) = \frac{1}{2} a_BR(t) = \frac{1}{2}a_B\, \mbox{sech}^2(\Omega_0 t + \delta)$, which may be realized by tuning the external magnetic field as 
\begin{equation}
B(t)=B_0+ \frac{ a_s^0 \,\,\, \Delta}{a_s^0 - \frac{1}{2} a_B \mbox{sech}^2(\Omega_0 t+\delta)}.
\end{equation}%
We point out here that such a form of scattering length has been realized in $^{7}$Li and $^{85}$Rb atoms~\cite{Strecker2002, Khaykovich2002, Courteille1998}. Figure \ref{fig2}(a) shows the  dark-bright one soliton solution for the above gain term where the amplitude of the wave packet remains constant. On the other hand, if we choose a periodic gain term, $\gamma(t) = \Omega' \sin( \Omega't)$, the amplitude shows oscillatory behaviour leading to periodic collapse and revival phenomena, see figure \ref{fig2}(b).
\begin{figure}[!ht]
\begin{center}
\includegraphics[width=0.8\linewidth]{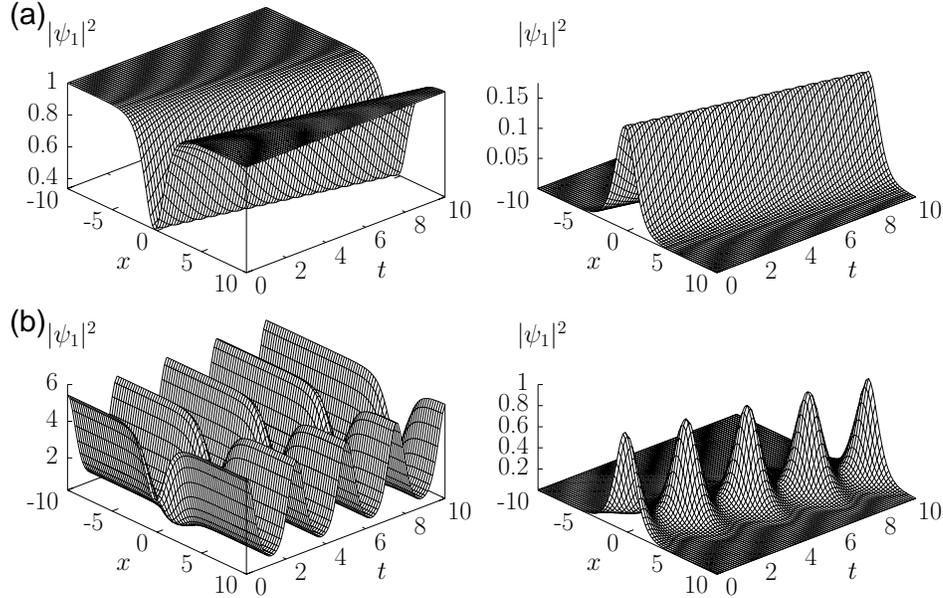}
\caption{Dynamics of dark (left) and bright (right) components of the one soliton in an expulsive trap potential for (a) $\gamma = \Omega_0 \tanh(\Omega_0 t)$ and (b)  $\gamma = \Omega' \sin(\Omega' t)$. The parameters are fixed at $a=0.7$, $b=0.5$,  $r_0=1/\sqrt{2}$, $c_1=c=0$, $\Omega' = 2.5$ and $\Omega_0=0.001$.}
\label{fig2}
\end{center}
\end{figure}
\begin{figure}[!ht]
\begin{center}
\includegraphics[width=0.6\linewidth]{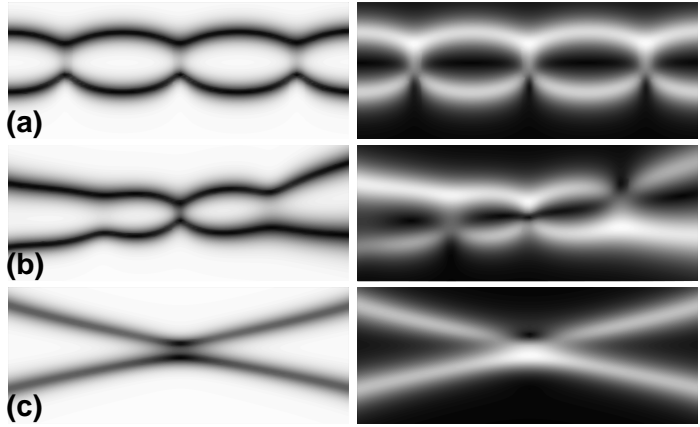}
\caption{Contour plots of two soliton interactions showing dark (left) and bright (right) components in the presence of expulsive trap potential. (a) Strong beating of two soliton bound states without interaction for $a_1=0.7$, $b_1=0$, $a_2=0.77$, $b_2=0$,  $\Omega_0=0.001$, (b) soliton interaction with beating for $a_1=0.7$, $b_1=-0.01$, $a_2=0.9$, $b_2=0.03$, $\Omega_0=0.01$, and (c) elastic collision of two solitons for  $a_1=0.7$, $b_1=0.3$, $a_2=0.7$, $b_2=-0.3$, $\Omega_0=0.05$. Here $x$ axis represents time in ms with $t \in (0, 200)$ for (a), $t \in (0, 300)$ for (b), and $t \in (0, 40)$ for (c). $y$ axis corresponds to the axial coordinate in $\mu$m, with $x \in (-6, 10)$ for (a) and (b), and $x \in (-10, 10)$ for (c). The other parameters are $r_0=1/\sqrt{2}$ and $c_1=c=0$.}
\label{fig3}
\end{center}
\end{figure}

We have also investigated different types of two soliton interactions for the gain term  $\gamma(t)=\Omega_0 \mbox{tanh}[\Omega_0 t]$ by suitably choosing the parameters $a_1$, $b_1$, $a_2$, $b_2$ and $\Omega_0$ in equations (\ref{sol2:q1}) and (\ref{sol2:q2}). Figure \ref{fig3}(a) shows  the interaction of dark and bright two solitons with strong beating effect without interaction while figure \ref{fig3}(b) illustrates interaction of two solitons with beating effect. Figure \ref{fig3}(c) shows the soliton interactions without beating. These type of soliton interactions are well known in the fiber optics context~\cite{Sheppard1997}.

\paragraph{Periodically modulated potential:-}  If we choose  $R(t) =1+ \omega \cos(\omega t+\delta) $ and  $ \gamma(t)=\displaystyle \frac{ \omega^2\sin(\omega t+ \delta)}{2[1+  \omega \cos(\omega t+\delta)]}$, $\omega < 1$ we get $\tilde{R}(t) = \sqrt{1+ \omega \cos(\omega t+\delta)}$ and the integrability condition (\ref{riccati}) gives $\displaystyle \Omega^2(t)= \omega^2\left\{1-\frac{4+ 10  \omega \cos(\omega t+\delta)+ 3 \omega^2 [1+\cos^2(\omega t+\delta)]}{4 \left[1+ \omega\cos(\omega t+\delta)\right]^2 }\right\}$ which is a periodically modulated trap. For the above case, we sketch the gain $\Gamma(t)=  \displaystyle  \frac{ \omega_\perp \omega^2 \sin(\omega t+ \delta)}{2[1+  \omega\cos(\omega t+\delta)]}$ and the corresponding choice of atomic scattering length $a_s(t)=\frac{a_B}{2}\left[1+ \omega\cos(\omega t+\delta)\right] $ in figure~\ref{fig1}(b), which may be realized by periodically tuning the external magnetic field as        
\begin{figure}[!ht]
\begin{center}
\includegraphics[width=0.8\linewidth]{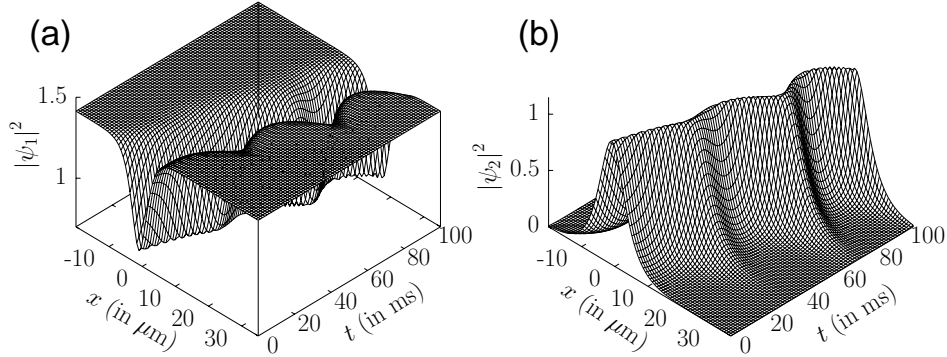}
\caption{Dynamics of dark component (a) and bright component (b) of snake-like one soliton in the presence of periodic modulated harmonic potential. The parameters are $a=0.3$, $b=0.2$, $r_0=1$, $c_1=c=0$, $\delta=0$ and $\Omega_0=0.4$.}
\label{fig4}
\end{center}
\end{figure}
\begin{equation}
B(t)=B_0+ \frac{a_s^0 \,\,\,\Delta}{a_s^0-\frac{1}{2} a_B \left[1+ \omega\cos(\omega t+\delta)\right] }.
\end{equation}
Figure \ref{fig4}(a) shows the snake-like effect of the one soliton solution for periodically modulated trap potential with $R(t) = \left[1+ \omega\cos(\omega t+\delta)\right] $ and  $ \gamma(t)=\displaystyle \frac{ \omega^2 \sin(\omega t+ \delta)}{2+ 2 \omega\cos(\omega t+\delta)}$. Similar snake effect has been demonstrated recently in scalar coupled nonautonomous NLS equations in BEC~\cite{Rajendran2008} and in the context of optical solitons~\cite{serkin2000}. Next we analyse different types of two soliton interactions for the periodically modulated potential and suitable choice of other parameters. Figure \ref{fig5}(a) shows the dark and bright non-interacting moving two solitons again with strong beating effect in bound state, where the velocities are zero ($b_1=b_2=0$) for $a_1=0.7$, $a_2=0.8$. Figure \ref{fig5}(b) shows non-interacting moving two solitons again with strong beating effect in unbound state for $a_1=0.7$, $a_2=0.8$, $b_1=b_2=0.1$. Figure \ref{fig5}(c) depicts the interacting two solitons with beating effect for $ a_1=0.7, a_2=0.85, b_1=0.1, b_2=0.12$. Finally figure \ref{fig5}(d) depicts the two soliton interaction without beating effect for $ a_1=0.7, a_2=0.85, b_1=0.1, b_2=0.17$. 
\begin{figure}[!ht]
\begin{center}
\includegraphics[width=0.6\linewidth]{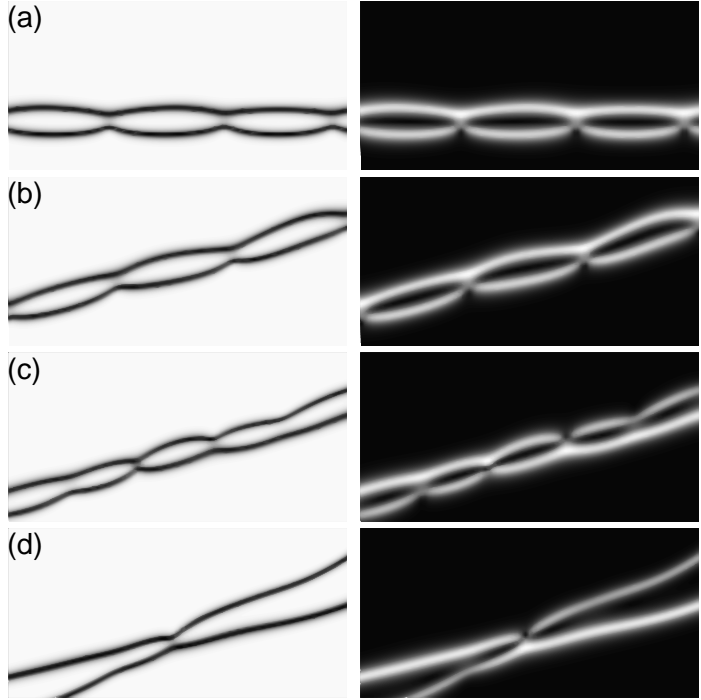}
\caption{Contour plots of matter wave soliton interactions of dark (left) and bright (right) components in a periodically modulated trap potential (a) noninteracting two soliton with strong beating effect in bound state for $a_1=0.7$, $a_2=0.8$, $b_1=b_2=0$, (b) noninteracting two soliton with strong beating effect in bound state for $a_1=0.7$, $a_2=0.8$, $b_1=b_2=0.1$, (c) two soliton interactions with beating effect for $a_1=0.7$, $a_2=0.85$, $b_1=0.1$, $b_2=0.12$, (d) two soliton interactions without beating effect for $a_1=0.7$, $a_2=0.85$, $b_1=0.1$, $b_2=0.17$. The other parameters are $r_0=1$, $c_1=c=0$, $\omega=0.1$ and $\delta = -5.0$. Here $x$ axis represents time in ms with $t \in (0, 300)$ and $y$ axis corresponds to the axial coordinate in $\mu$m, with $x \in (-10, 30)$.}
\label{fig5}
\end{center}
\end{figure}

Recent experimental studies show the above types of dark and dark-bright soliton oscillations and interactions in single and two component BECs~\cite{Christoph2008, Weller2008, Stellmer2008}. However these experiments do not consider the interaction of thermal cloud or time dependent interatomic interaction. Our above studies clearly suggest the possibility of observing them in experiments.

In summary, we have investigated the exact dark-bright one and two soliton solutions of the two-component BECs with time varying parameters such as s-wave scattering length and gain/loss term. On mapping the two coupled GP equation onto the coupled NLS equation under certain conditions, we have deduced different kinds of dark-bright one soliton solutions and interaction of two solitons for time independent and periodically modulated trap potentials. The present study provides an understanding of the possible mechanism for soliton excitations in multi-component BECs. These excitations can be realized in experiments by suitable control of time dependent trap parameters, atomic interaction and interaction with thermal cloud.

\ack

This work is supported in part by Department of Science and Technology (DST), Government of India - DST-IRHPA project (SR and ML), and DST Ramanna Fellowship (ML). The work of PM forms part of DST-FTYS project (Ref. No. SR/FTP/PS-79/2005).

\section*{References}


\providecommand{\newblock}{}

\end{document}